\documentclass[11pt,a4paper,english,amssymb,nofootinbib,superscriptaddress]{revtex4}
\usepackage{lmodern}

\usepackage[T1]{fontenc}
\usepackage[latin9]{inputenc}
\usepackage{babel}
\usepackage{amsmath}
\usepackage{graphicx}
\usepackage{amssymb}
\usepackage{esint}
\usepackage[unicode=true, pdfusetitle,
bookmarks=true,bookmarksnumbered=false,bookmarksopen=false,
breaklinks=false,pdfborder={0 0 1},backref=false,colorlinks=false]
{hyperref}
\def\b{\begin{equation}}
\def\e{\end{equation}}

\makeatletter
\@ifundefined{textcolor}{}
{%
 \definecolor{BLACK}{gray}{0}
 \definecolor{WHITE}{gray}{1}
 \definecolor{RED}{rgb}{1,0,0}
 \definecolor{GREEN}{rgb}{0,1,0}
 \definecolor{BLUE}{rgb}{0,0,1}
 \definecolor{CYAN}{cmyk}{1,0,0,0}
 \definecolor{MAGENTA}{cmyk}{0,1,0,0}
 \definecolor{YELLOW}{cmyk}{0,0,1,0}
 }

\usepackage{latexsym}\usepackage{bm}

\makeatother

\makeatother

\begin{document}
\title{Asymptotically flat black holes in 2 + 1 dimensions }

\author{G{\"{o}}khan Alka\c{c}}

\affiliation{Van Swinderen Institute for Particle Physics and Gravity,\\ University of Groningen, Nijenborgh 4, 9747 AG Groningen, The Netherlands,}
\email{g.alkac@rug.nl, kercan@metu.edu.tr, btekin@metu.edu.tr}

\author{Ercan Kilicarslan}

\affiliation{Van Swinderen Institute for Particle Physics and Gravity,\\ University of Groningen, Nijenborgh 4, 9747 AG Groningen, The Netherlands,}

\affiliation{Department of Physics,\\
 Middle East Technical University, 06800, Ankara, Turkey}

\author{Bayram Tekin}

\affiliation{Department of Physics,\\
 Middle East Technical University, 06800, Ankara, Turkey}
 
\date{\today}

\begin{abstract}
Asymptotically flat black holes in $2+1$ dimensions are a rarity. We study the recently found black flower solutions (asymptotically flat black holes with deformed horizons), static black holes, rotating black holes and  the dynamical black flowers (black holes with radiative gravitons ) of the purely quadratic version of new massive gravity. We show how they appear in this theory and we also show that they are also solutions to the infinite order extended version of the new massive gravity, that is the Born-Infeld extension of new massive gravity with an amputated Einsteinian piece.  The  same metrics also solve the topologically extended versions of these theories, with modified conserved charges and the thermodynamical quantities, such as the Wald entropy.  Besides these we find new conformally flat radiating type  solutions to these extended gravity models. We also show that these metrics do not arise in Einstein's gravity coupled to physical perfect fluids.

\end{abstract}

\maketitle

\section{Introduction}

In his 1884 book titled  "{\it Flatland: A Romance of Many Dimensions}",  Edwin A. Abbott 's witty narrator, " the Square", who  longs for and theorizes about the extra dimensions beyond his World of $2+1$, complains about the dullness of life in such a universe as  "{\it How can it be otherwise, when all one's prospect, all one's landscapes, historical pieces, portraits, flowers, still life, are nothing but a single line, with no varieties except degrees of brightness and obscurity? }"  From today's  vantage point, in the centenary of the birth of general relativity, we know that the state of affairs in flatland is not that bleak if you do not ignore gravity. (Note that Abbott did mention the existence of  gravity in Flatland, as the word "gravitation" appears in one place just like the word flowers.)   First there is the famous BTZ \cite{BTZ} black hole in Einstein's theory with a negative cosmological constant. But this black hole is locally anti-de Sitter space (AdS), 
and globally a rather complicated identification of AdS and hence, a local observer, such as the "the Square", who does not have access to the major part of the $2+1$ dimensional universe will not notice this black hole (or more properly, will not know that he is living in a black hole spacetime). Since in 3 dimensions, the Ricci tensor and the Riemann tensor are double-duals of each other, there is no local gravity in Einstein's theory save the regions with sources. On the other hand, moving beyond Einstein's theory,  one realizes that there are asymptotically flat 
black holes, with nontrivial, nonconstant scalar curvature and other curvature invariants and so these black holes actually create tidal forces  in nearby objects. So in principle they can be "observed"  by the flat-beings, in the  same way as the more realistic ones in three spatial dimensions. Of course the rationale to study gravity in a lower dimensional setting is to possibly learn something about the inner-workings of the strong regime of gravity and hopefully quantum gravity  in our universe.   For this purpose black holes are the lamp-posts for any theory of gravity: the plethora of the work done on the BTZ black hole and its variants and AdS/CFT correspondence is a testament to this. The new black holes, the black flowers and their static, rotating and radiating versions deserve a greater scrutiny as they also might hide possible hints about quantum gravity.  Of course, being asymptotically flat, they are not canonical objects to look at in the AdS/CFT context. So, holography in flat spacetime needs to be better understood. 

In this work, we show in more clear terms how they arise as solutions to the purely quadratic part of new massive gravity (NMG) \cite{NMG}. This purely quadratic theory was  studied in \cite{KGravity} as a ghost-free theory of massless gravitons in 3D. (For the sake of simplification of the names of the theories, as there will be several of those, we shall call this theory $K$- gravity, $K$- being the tuned quadratic curvature invariant that we shall write.) Utilizing the recent observation \cite{MMG2,Altas} that the field equations of ${K}$-gravity split into two natural parts as  $K_{\mu\nu} = J_{\mu \nu} + H_{\mu \nu}$, whose explicit forms will be given below, where the $H$-tensor is the 3 dimensional version of the Bach tensor that vanishes for conformally flat (and conformally Einstein) geometries. On the other hand the $J$-tensor does not have any derivatives of the curvature  and appeared as an interesting, bulk and boundary unitary deformation of topologically massive gravity (TMG) \cite{DJT} to the minimal massive gravity (MMG) \cite{mmg}.  All the types of the black flower solution will appear as a conformally flat solution ($ H_
 {\mu \nu}=0$),  that also has vanishing $J$-tensor. This realization and dissection of the $K$-gravity's equations  are crucial to upgrade the solution to other extended theories of gravity, such as the Born-Infeld extension of NMG, without the Einstein term ( which we shall call BIK-gravity) and their Chern-Simons modified versions. We will also be able to show that there are in fact 3 types of such solutions and no more: Type-$D$ solutions to which the black flower, static and rotating metrics belong to, Type-II solutions to which the dynamical black flower is an example of  and Type-${N}$ solutions.

\section{Black Holes in Purely quadratic theory: K-gravity}

This was already done in \cite{Barnich, Tempo, Fareghbal} partly inspired from earlier solutions \cite{Oliva,Oliva2,Blagojevic} but as stated above, we shall make the appearance of the solution more transparent to be able to upgrade it to other extended theories, classify the solutions and show that there are no other possible classes under the given conditions. The action of the $K$-gravity  is 
\begin{eqnarray}
 I= \frac{1}{16 \pi G}\int  d^3 x \, \sqrt{-g}\, K,    \hskip 1 cm K \equiv  R_{\mu \nu} R^{\mu \nu} - \frac{3}{8}R^2, 
\label{action}
\end{eqnarray}
where the 3D Newton's constant $G$ has dimensions of mass which makes the theory power-counting super-renormalizable.  The theory has a massless ghost-free graviton about its unique maximally symmetric, flat vacuum. The source-free field equations   ($K_{\mu \nu} =0$) read
\begin{equation}
2\square R_{\mu\nu}-\frac{1}{2}\nabla_\mu\nabla_\nu R-\frac{1}{2}g_{\mu\nu}\square R+4R_{\mu\rho\nu\sigma}R^{\rho\sigma}-g_{\mu\nu}R_{\rho\sigma}R^{\rho\sigma}
-\frac{3}{2}RR_{\mu\nu}+\frac{3}{8}g_{\mu\nu}R^2=0.
\end{equation}
They look somewhat cumbersome but as observed in \cite{MMG2}, these field equations split into two natural pieces
\begin{equation}
 K_{\mu \nu}=  J_{ \mu \nu} + H_{\mu \nu}=0,
\end{equation}
where  the $J$ and  $H$-tensors are defined as  
\begin{equation}
 J^{\mu\nu}\equiv \frac{1}{2}\eta^{\mu\rho\sigma}\eta^{\nu\tau\alpha}S_{\rho\tau}S_{\sigma\alpha}, \hskip 1cm   H_{ \mu \nu} \equiv  \frac{1}{2}\eta_{\mu}\,^{ \alpha\beta}\nabla_{\alpha} C_{\beta \nu} + \frac{1}{2}\eta_{\nu}\,^{ \alpha\beta}\nabla_{\alpha} C_{\beta \mu},
\end{equation}
with $\eta^{\nu\tau\eta}$ being the completely antisymmetric tensor, $S_{\mu \nu} = R_{\mu \nu} -\frac{1}{4} g_{\mu\nu}R$ the Schouten tensor and $C_{\mu\nu}=\eta_{\mu}\thinspace^{\alpha\beta}\nabla_{\alpha}S_{\beta \nu}$ the Cotton tensor.   Observe that $H_{\mu\nu}$ is traceless and so  the trace of the field equations yield
$K = g^{\mu \nu} K_{\mu\nu} =  g^{\mu \nu} J_{\mu\nu}$. As for the covariant divergence of these tensors one has: $ \nabla_\mu H^{\mu \nu} = -  \nabla_\mu J^{\mu \nu} =  \eta^{\nu\alpha\beta}S_{\alpha \sigma} C_{\beta}\,^\sigma$. As we shall see, this compact form of writing the equations is not merely about aesthetics, but it will help us understand the solution better since we shall know which parts of the equations vanish on their own.

It is clear from the above discussion that {\it all} the solutions of $K$-gravity satisfy the on-shell vanishing of $K$- (and so the action)
\begin{equation}
R_{\mu \nu} R^{\mu \nu} = \frac{3}{8}R^2,
\end{equation}
which can be recast  as
\begin{equation}
I_1 \equiv   \tilde{R}_\mu^\nu \tilde{R}_\nu^\mu  = \frac{1}{24}R^2,
\label{I1eqn}
\end{equation}
where $\tilde{R}_{\mu\nu}$ is the traceless Ricci tensor and $I_1$ is a curvature invariant which we shall use to understand the Segre classification (based on the traceless Ricci tensor) of the solutions along with $R$ and 
a third one:
\begin{equation}
 I_2 \equiv \tilde{R}_\mu^\nu \tilde{R}_\nu^\rho  \tilde{R}_\rho^\mu,
\end{equation} 
which form the  algebraically independent set for a 3 dimensional symmetric tensor as is clear form the Cayley-Hamilton theorem and Schouten identities applied to the $3\times 3$  matrix $(\tilde{R}_\nu^\mu)$. Now consider {\it all}  the solutions of the theory with $H_{\mu \nu} =0$, and hence one must set  $J_{\mu \nu}=0$ as well yielding
\begin{equation}
J_{\mu \nu} =
-\widetilde{R}_{\mu \rho}\widetilde{R}^{\rho}_\nu+\frac{1}{3}g_{\mu\nu}I_1+\frac{1}{12}R \widetilde{R}_{\mu\nu} =0,
\label{traceless_J}
\end{equation}where we wrote the explicit form of this tensor and used the fact that  its trace must vanish. At this stage the solutions bifurcate into two main classes: $R \ne 0$ and $R=0$.
Let us consider these cases separately as they will lead to spaces of different types.

\subsection{$ R \ne 0$ case }

For this case  the traceless Ricci tensor can be expressed from (\ref{traceless_J}) as  
\begin{equation}
\widetilde{R}_{\mu\nu} = \frac{12}{R} \Big ( \widetilde{R}_{\mu \rho}\widetilde{R}^{\rho}_\nu-\frac{1}{3}g_{\mu\nu}I_1 \Big ).
\end{equation}
Multiplying this with $\widetilde{R}^{\mu\nu}$, one finds that the 3 curvature invariants are related as 
\begin{equation}
I_1 = \frac{12}{R}I_2, \hskip 1 cm I_2 = \frac{R^3}{288}, \hskip 1 cm I_1^3 = 6 I_2^2.   
\end{equation}
Hence according to the Segre classification, all such solutions of this theory are either  type-$D$ (metrics for localized objects) or type-$II$ (metrics for localized objects with gravitational radiation). Of course, this analysis does not yet tell us that there are solutions to the theory. We must search for them explicitly.  The best way to start with is a circularly symmetric ansatz:
\begin{equation}
ds^2=- f(r)dt^2+\frac{dr^2}{f(r)}+r^2d\phi^2,
\end{equation}
Computing $H_{\mu \nu}=0$, one finds that  the metric function satisfies  $f(r) = b r-\mu + c r^2$, which yields the most general spherically symmetric and conformally flat metric.  Inserting this solution to the  remaining piece of the equations,  $J_{\mu \nu} =0$ equation, one finds that  $c=0$ is required. 
Hence the following, conformally flat, static and asymptotically locally flat metric  
\begin{equation}
ds^2=-(br-\mu)dt^2+\frac{dr^2}{br-\mu}+r^2d\phi^2,
\label{static}
\end{equation}
is a solution of $K$-gravity  with two constants $b$ and $\mu$. It is a static, asymptotically flat black hole as discussed in \cite{Oliva, Barnich,Tempo} with a circular Schwarzschild horizon at   $r_s= \frac{{\mu}}{b}$  as long as $\mu >0$ and $b >0$.  One can easily see that its traceless  Ricci tensor is of Type-$D_s$ given as 
\begin{equation}
 \tilde{R}_{\mu\nu} = -\frac{R}{12} \Big ( g_{\mu \nu} - 3 \xi_\mu \xi_\nu \Big ),
\end{equation}
where the scalar curvature and the  relevant vector are given as 
\begin{equation}
R =-\frac{2 b}{r}, \hskip 1 cm \xi_\mu =- r\delta_\mu^\phi,
\end{equation}
with $ \xi_\mu \xi^\mu = 1 $.  The fact that $\xi_\mu$ is a space-like vector (hence the subscript $s$ in $D_s$) is important as we shall note below.  Note also that  the scalar curvature $R$ and the invariants $I_1$, $I_2$ blow up at the singularity $r=0$, hidden safely inside the Schwarzschild horizon $r_s$. Going to the Euclidean version, near the event horizon one finds that the Hawking temperature of the black hole is 
$T = \frac{b}{4 \pi}$. Hence the dimensional parameter $b$ is related to the mass of the black hole while the dimensionless parameter $\mu$ is a gravitational hair \cite{Barnich}.

Other solutions can be generated from the above static black hole. After a transformation ($u= t-r^*$ with $dr^* = \frac{dr} { b r -\mu}$ ) described in \cite{Barnich},  the static metric becomes
\begin{equation}
ds^2=-(br-\mu)du^2-2 du dr+r^2d\phi^2.
\end{equation}
With this form of the metric, one can introduce a deformation [$h(u,\phi)$] along the spacelike Killing vector $\partial_\phi$ as
\begin{equation}
ds^2=-(br-\mu)du^2-2 du dr+(r-h(u,\phi))^2d\phi^2.
\end{equation}
Inserting this ansatz to $H_{\mu \nu}=0$, one finds that the following linear equation must be satisfied\footnote{Note that one can actually take a  more general metric such as $ds^2=-g(r)du^2-2 du dr+(r-h(u,\phi))^2d\phi^2$ and find solutions to  $H_{\mu \nu} =0$ equations, but, it turns out unless  $g(r) =br-\mu$, one does not have $J_{\mu \nu}=0$.  }
\begin{equation}
\partial_u(\partial_u+\frac{b}{2}) h=0,
\end{equation}
whose solution is 
\begin{equation}
h(u,\phi)=A(\phi)+B(\phi) e^{-\frac{b}{2} u},
\end{equation}
where $A(\phi)$ and $B(\phi)$  are arbitrary periodic functions of  the angular coordinate $\phi$. As a simple example of a black flower see Fig.1. For this solution  one has $J_{\mu \nu}=0$ automatically  and the equations of ${K}$-gravity are solved. Let us rewrite the metric in terms of the original coordinates to get a better picture of its nature: 
\begin{equation}
ds^2=-(br-\mu)dt^2 +\frac{dr^2}{br-\mu}+\Big(r- h(t,r,\phi) \Big)^2d\phi^2,
\label{flo_met}
\end{equation}
where we have 
\begin{equation}
h(t,r,\phi) = A(\phi)+B(\phi) e^{-\frac{b}{2} t}( b r - \mu)^{1/2}.
\end{equation}
This solution is again Type-$D_s$  or Type-$II$ with the spacelike vector reading as
\begin{equation}
\xi_{\mu} = \Big( h(t,r,\phi)-r\Big ) \delta_\mu^\phi.
\end{equation}
When $B(\phi)$=0, time dependence drops out and one has a static, circularly nonsymmetric black hole (black flower) which is Type-$D_s$. But when $B(\phi) \ne 0$, then the  spacetime is dynamical and the black flower is formed only at  $t \rightarrow \infty$ for b>0.  Namely, one can interpret  either as a black hole radiating gravitons or as in-falling self-gravitating gravitons forming a black flower in the far future.

So the full metric is
\begin{equation}
g_{\mu \nu} = g_{\mu \nu}^S+ \xi_\mu \xi_\nu,
\end{equation}
where   $g_{\mu \nu}^S$ refers to the static circularly symmetric solution.  This form suggests that one can write it in a double Kerr-Schild way as  
\begin{equation}
g_{\mu \nu} = \eta_{\mu \nu}+ f(r)\lambda_\mu \lambda_\nu+  \xi_\mu \xi_\nu,
\end{equation}
with $\eta_{\mu  \nu} $ the flat metric in the $u,r,\phi$ coordinates and  $f(r) = br -\mu$, $\lambda_\mu = \delta_\mu^u$.  While formally, this is correct, since the $\xi$-vector is not null, quick exact  linearization techniques of Kerr-Schild metrics do not follow \cite{Gurses_Gursey}.

For completeness let us note that  in the $u,r,\phi$ coordinates one simply has 
\begin{equation}
\xi_\mu =\Big( h(u,\phi)-r\Big ) \delta_\mu^\phi,
\end{equation}
which has vanishing divergence $\nabla_\mu \xi^\mu=0$, but it is not a Killing vector. Let us note that for both static asymptotically flat black hole and the black flower,  and the dynamical black flower,  the Einstein tensor reads as
\begin{equation}
G_{\mu \nu} = -\frac{R}{4} \Big  ( g_{\mu \nu} - \xi_\mu \xi_\nu  \Big ).
\end{equation}
Defining the right-hand side as the energy-momentum tensor $T_{\mu \nu}$ one has
\begin{equation}
T_{\mu \nu}=-\frac{R}{4} \Big  ( g_{\mu \nu} - \xi_\mu \xi_\nu  \Big ),
\end{equation}
and one might wonder if these solutions appear in Einstein's gravity coupled to perfect fluid \cite{Gurses1,Gurses2} with the pressure ($P$) and mass density ($\rho$) given as 
$P =- \frac{R}{4}$,  $\rho = \frac{R}{2}$. While {\it formally} this is a strong energy condition-violating fluid, since the fluid velocity $\xi_\mu$ is spacelike these solutions do not exist in Einstein's theory coupled to perfect fluids. Before we move on to the other case, let us note that for the black flower, the scalar curvature is
\begin{equation}
R = -\frac{2 b} {r-h(u,\phi)}.
\label{scalar_R}
\end{equation} 
In the figure, we have depicted an example of a  simple black flower $[B(\phi)=0.]$  One must be careful though in the interpretation: from (\ref{scalar_R}), it is clear that the scalar curvature is singular for the values of the radial coordinate $r= A(\phi)$  and hence one must restrict $ r > A(\phi)$. On the other  hand,  it is also clear from (\ref{flo_met}) that the event horizon is at $r_s = \frac{b}{\mu}$. If the singularity is pushed to the origin of the coordinates, than the horizon takes the noncircular shape. Or another way to see is to look at the induced metric on the horizon 
$g_{\phi \phi} = (r_s - A(\phi))^2$ as suggested in \cite{Barnich}.
\begin{figure}[h]
\centering
\includegraphics[width=0.5\textwidth]{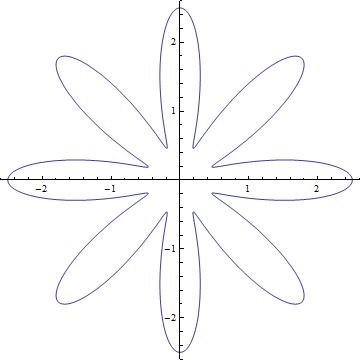}
\caption{ A depiction of the black flower for the specific choice of unitless  
$A(\phi) = \cos( 8 \phi) + 1.5$ .}
\label{hyp_embed}
\end{figure}
\newpage
{\bf  2) $ R = 0$ case }

For this case one has
\begin{equation}
\frac{1}{3}g_{\mu\nu}I_1=\widetilde{R}_{\mu \rho}\widetilde{R}^{\rho}_\nu.
\end{equation}
Multiplying this with $\widetilde{R}^{\mu\nu}$, one arrives at 
\begin{equation}
I_2=0.
\end{equation}
also from (\ref{I1eqn}) 
\begin{equation}
I_1=0.  
\end{equation}
Hence according to the Segre classification, all such solutions are
Type-$N_s$ with vanishing curvature invariants
\begin{equation}
 R_{\mu\nu} =  \xi_\mu \xi_\nu,
\end{equation}
where $\xi_\mu$ is a null-like vector. For this type of solutions, we can give a new example as 
\begin{equation}
ds^2=F^2(t,x,y)(-dt^2+dx^2+dy^2),
\end{equation} 
where $F^2(t,x,y)=(at+bx+cy)^q$ with $b=\pm \sqrt{a^2+c^2}$ where $q$ is an arbitrary real number.

\section{Born-Infeld extension of K-Gravity}
Let us consider the Born-Infeld extension of $K$-gravity given by the action \cite{gullu,gullu2,Tahsin_tez}
\begin{equation}
 I=-\frac{\gamma^{2}}{4 \pi G}\int d^3x\bigg[\sqrt{-\det \Big( g_{\mu\nu}+\frac{\sigma}{\gamma}G_{\mu\nu} \Big )}-(1-\frac{\sigma}{4\gamma}R)\sqrt{-g}\bigg],
 \label{BIKaction}
\end{equation}
where $\gamma$ is the BI parameter with  $[ \gamma ] = mass^2$ dimensions. This parameter is needed to be able to define a BI extension of gravity. It controls the scale where the higher derivative terms significantly change the structure of the theory. At the lowest order in the curvature expansion, this BI action reproduces that of $K$-gravity and $\gamma$ plays no role. But in general there are two scales: the 3D Planck scale $M_p= 16 \pi G$ and  $\sqrt{\gamma}$. Of course one expects the latter to be several orders of magnitude smaller than the former. The theory has a unique maximally symmetric vacuum, that is the flat space and a massless unitary graviton about the vacuum. To find field equations of (\ref{BIKaction}), we define a new tensor as $M_{\mu\nu}=g_{\mu\nu}+\frac{\sigma}{\gamma}G_{\mu\nu}$. Now, (\ref{BIKaction}) can be written as 
\begin{equation}
 I=-\frac{\gamma^{2}}{4 \pi G}\int d^3x\bigg[\sqrt{-M}-(1-\frac{\sigma}{4\gamma}R)\sqrt{-g}\bigg].
\end{equation}
Let us focus on the variation of first part of the action;
\begin{equation}
\begin{aligned}
& I_M=\int d^3x \sqrt{-M}\\
&\delta I_M=\frac{1}{2}\int d^3x\sqrt{-M}\,\sum_{\mu,\nu}(M^{-1})_{\mu\nu}\,\delta M_{\mu\nu}\\
&\delta I_M=\frac{1}{2}\int d^3x\sqrt{-g}\,\bigg(\sqrt{\frac{M}{g}}\sum_{\mu,\nu}(M^{-1})_{\mu\nu}\,\delta M_{\mu\nu}\bigg)\\
&\delta I_M=\frac{1}{2}\int d^3x\sqrt{-g}\,B^{\mu\nu}\delta M_{\mu\nu},
\end{aligned}
\end{equation}
where the B tensor is defined by $ B^{\mu\nu}=\sqrt{\frac{M}{g}}(M^{-1})_{\mu\nu}$ and $(M^{-1})_{\mu\nu}$ are the elements of the inverse of the matrix $M_{\mu\nu}$.  Using this form is particularly useful in finding field equations, which leads to
\begin{equation}
\begin{aligned}
&-B_{\mu\nu}+g_{\mu\nu}+\frac{\sigma}{\gamma}\bigg(-\nabla_\alpha\nabla_({_\mu} B_\nu{_)}^\alpha
+\frac{1}{2}(\nabla_\alpha\nabla_\beta B^{\alpha\beta}-\square B)g_{\mu\nu}\\&+\frac{1}{2}\square B_{\mu\nu}+\frac{1}{2}B_{\mu\nu}R+\frac{1}{2}\nabla_\mu\nabla_\nu B
-\frac{1}{2}B R_{\mu\nu}+\frac{1}{2}G_{\mu\nu}\bigg)=0,
\end{aligned}
\end{equation}
where $B=g_{\mu\nu}B^{\mu\nu}$. See also \cite{Nam} for a similar form of these equations.
The metrics  (\ref{static}) and (\ref{flo_met}) solve the BIK equations without a change of the metric functions. Let us rewrite the Born-Infeld extension of $K$-gravity given by the action  
\begin{equation}
 I_{BIK}=-\frac{\gamma^{2}}{4 \pi G}\int d^3x\sqrt{-g}\bigg(F(R,S_1,S_2)+\frac{\sigma}{4\gamma}R\bigg),
 \label{QGaction}
\end{equation}
where 
\begin{equation}
\begin{aligned}
 & F(R,S_1,S_2)=\sqrt{1-\frac{\sigma}{2\gamma}(R+\frac{\sigma}{\gamma}S_1-\frac{1}{12\gamma^2}S_2)}-1,\\
 & S_1=R^2_{\alpha\beta}-\frac{R^2}{2}, \hskip 0.5cm S_2=8 R^{\alpha\beta}R_{\alpha\sigma}R^\sigma\,_{\beta}-6RR^2_{\alpha\beta}+R^3.
 \end{aligned}
\end{equation}
We shall need these in the computation of the entropy and the mass of the black holes.

\subsection{Some useful formulas for the calculation of entropy and mass }

What we present here is common knowledge, hence we do not go into details but just quote the final results. We will calculate the geometrical entropy using the Wald formula where the black hole entropy is computed to be the Noether charge on the horizon \cite{Wald,Wald2,Wald3}. There exists a simplified form of this formula given by \cite{Jacobson,Myers,Saida}
\begin{equation}
S=-2\pi \int d\phi\sqrt{g_{\phi\phi}}\frac{\partial \cal{L}}{\partial R_{\mu\nu}}\epsilon_\mu\,^\alpha\epsilon_{\nu\alpha}.
\end{equation}
Here, the binormal $\epsilon_{\nu\beta}$ is defined through the time-like killing vector $\chi^\mu=(-1,0,-\Omega) $, with $\Omega=-\frac{g_{t\phi}}{g_{\phi\phi}}$ being the angular velocity of the event horizon.   One has the relation
\begin{equation}
\epsilon_{\mu\nu}=\frac{1}{\kappa}\nabla_\mu\chi_\nu,
\end{equation}
where $\kappa$ is the surface gravity defined as
\begin{equation}
\kappa=\sqrt{-\frac{1}{2}\nabla^\mu\chi^\nu\nabla_\mu\chi_\nu}.
\end{equation}

In the presence of the Chern-Simons term which we will define below  in (\ref{CS-K}), the black hole entropy is given in \cite{CSentropy,CSentropy2}
\begin{equation}
S=\frac{2\pi}{\mu_{CS}} \int dx^\sigma\Gamma^\nu_{\mu\sigma}\epsilon^\mu\,_\nu\,.
\end{equation}
Using the first law of thermodynamics, we also compute the mass of the static and rotating asymptotically locally flat black holes. It is an interesting property of these solutions that the angular velocity of the horizon $\Omega$ vanishes and the first law becomes $TdS=dM$. As a result, there is no way to compute the angular momentum with the help of the first law. For the $K$-gravity, it was computed in \cite{Barnich}; however the angular momentum for the BIK theory remains to be computed by a direct method. 

To compute the entropy and the mass of black holes in BIK gravity, we need the following expression 

\begin{equation}
\frac{\partial F}{\partial R_{\mu\nu}}=\frac{-\sigma}{4\gamma (F+1)}\bigg([1-\frac{\sigma}{\gamma}R+\frac{1}{2\gamma^2}(R_{\rho\sigma}^2-\frac{1}{2}R^2)]g^{\mu\nu}+\frac{2}{\gamma}(\sigma+\frac{1}{2\gamma}R)R^{\mu\nu}-\frac{2}{\gamma^2}R_{\mu\rho}R^\rho\,_\nu\bigg).
\end{equation}
For the static black hole (\ref{static}) and the black flower (\ref{flo_met}), one finds through the above methods
\begin{equation}
S=\frac{\pi b }{4G}, \hskip 1 cm M=\frac{ b^2}{32G},
\label{ent_mass}
\end{equation}
which are exactly the same values found for the $K$-gravity \cite{Barnich}. $K$- gravity has a rotating solution  (\cite{Oliva,Giribet}) given as 
\begin{equation}
ds^2 = -N F du^2 - 2 N^{1/2} du dr + (r^2+ r_0^2) ( d\phi+ N^\phi du)^2,
\label{rotatingmetric}
\end{equation}
with
\begin{equation}
F= b r - \mu, \hskip 0.5 cm N = \frac{ ( 8 r + a^2 b)^2}{ 64(r^2 + r_0^2)}, \hskip 0.5cm N^\phi = - \frac{a}{2}\Big( \frac{ b r- \mu }{ r^2+ r_0^2} \Big), \hskip 0.5 cm r_0^2 = \frac{a^2}{4} \Big ( \mu + \frac{a^2 b^2}{16} \Big) .
\end{equation}
In \cite{Fareghbal}, this solution is obtained by taking the flat-space limit of NMG and a contracted conformal field theory with the BMS$_3$ symmetries is proposed as a dual description of K-gravity. This is a type-$D_s$ metric with the following properties
\begin{equation}
 \tilde{R}_{\mu\nu} = -\frac{R}{12} \Big ( g_{\mu \nu} - 3 \xi_\mu \xi_\nu \Big ), \hskip 1 cm \xi_\mu =\left\{0,-\frac{1}{2} a \sqrt{\frac{1}{r_0^2+r^2}},-\frac{a^2
   b}{8}-r\right\}, \hskip  1 cm \xi_\mu \xi^\mu =1.
\end{equation}
With this form of the Ricci tensor, one can show that the rotating metric, remarkably, also  solves the BIK gravity with  a lot more complicated field equations. 
Calculation of the entropy and the mass also yields the same result as the static and the black flower cases (\ref{ent_mass}).

Before we move on to the Chern-Simons extensions of these theories, let us note that  $K$-gravity admits a second Born-Infeld type extension with $M^{(2)}_{\mu\nu} = g_{\mu\nu}+\frac{\sigma}{\gamma}\left(R_{\mu\nu}- \frac{1}{6} g_{\mu\nu} R\right) $.

\subsection{Topologically extended K-gravity and BIK gravity}

The Lagrangian density of topologically extended K-gravity \cite{DJT}
 \begin{equation}
 {\cal L}= \sqrt{-g} \, \Bigg \{ \frac{1}{16 \pi G}( R_{\mu \nu} R^{\mu \nu} - \frac{3}{8}R^2) +\frac{1}{2 \mu_{CS}} \, \eta^{\mu \nu \alpha} \Gamma^\beta{_{\mu \sigma}} \Big (\partial_\nu \Gamma^\sigma{_{\alpha \beta}}+\frac{2}{3} \Gamma^\sigma{_{\nu \lambda}}  \Gamma^\lambda{_{\alpha \beta}} \Big ) \Bigg \},
 \label{CS-K}
\end{equation}
Let us calculate the mass and entropy of the static and rotating black holes.
\begin{enumerate}
\item 
The theory has an asymptotically flat black hole solution. Using the static and asymptotically locally flat metric (\ref{static}), one can obtain
\begin{equation}
S=\frac{\pi b }{4G}, \hskip 1 cm M=\frac{ b^2}{32G}.
\end{equation}
\item The theory has a rotating black hole solution. Using the rotating asymptotically locally flat metric (\ref{rotatingmetric}), one can obtain
\begin{equation}
S=\pi b \Big (\frac{1 }{4G}+\frac{2\pi a}{\mu_{CS}} \Big), \hskip 1 cm M=b^2 \Big (\frac{1 }{32G}+\frac{3\pi a}{4\mu_{CS}} \Big ).
\end{equation}
\end{enumerate}
The Lagrangian density of topologically extended BIK gravity
 \begin{equation}
 {\cal L}= \sqrt{-g} \, \Bigg \{ -\frac{\gamma^2}{4 \pi G}\bigg(F(R,K,S)+\frac{\sigma}{4 \gamma}R\bigg) +\frac{1}{ 2\mu_{CS}} \, \eta^{\mu \nu \alpha} \Gamma^\beta{_{\mu \sigma}} \Big (\partial_\nu \Gamma^\sigma{_{\alpha \beta}}+\frac{2}{3} \Gamma^\sigma{_{\nu \lambda}}  \Gamma^\lambda{_{\alpha \beta}} \Big ) \Bigg \},
\end{equation}
has the same entropy and mass expressions for these solutions. Hence the Chern-Simons term does its job and changes the thermodynamics and conserved charges of the rotating solution. Note that naively, for a choice of the Chern-Simons parameter $\mu_{CS}$, the mass or the entropy of the black hole vanishes.

\section{Conclusion}
In this work we have studied the recently found asymptotically flat black holes (static ones, rotating ones, the black flowers) and the dynamical space-times of purely quadratic part of the three dimensional new massive gravity. After identifying their types, according to the traceless Ricci tensor, we were able to upgrade these solutions to the Born-Infeld extension of the same theory which in principle has infinite powers of curvature. As expected, static black hole, rotating black hole and the black flower are type-$D$, albeit with a spacelike vector, hence do not appear as solutions to Einstein's theory coupled to perfect "physical fluids".  Dynamical black flower is of type-$II$  representing the appearance static black hole in the far future either forming due to the infalling gravitons  or outgoing gravitons from a black hole.  We also gave some new type-N solutions.

By dissecting the quadratic part of the field equations of new massive gravity into two tensors one of which vanishes for conformally Einstein spaces and one of which is purely algebraic in the powers of the curvature, one understands better how these rather rare curiosities arise in the claustrophobic theoretical lab of 2+1 dimensions.  The fact that these solutions are conformally flat allows us to extend them trivially to be solutions of Chern-Simons Lagrangian added versions of the same theories. Even though the solutions are intact, the conserved charges and the entropy get corrections from the Chern-Simons term which computed. An exhaustive reach for all solutions with nontrivial 3D Bach tensor (namely, $H_{\mu \nu} \ne 0$) also needs to be done in the $K$-gravity and the BIK gravity along the lines of  done for the new massive gravity and other extended 3D theories{\cite{Gurses1,Ahmedovprl,Ahmedovplb,Gurses:2011fv,Ahmedovprd}.

\section{\label{ackno} Acknowledgements}
The work of B.T. is supported by the TUBITAK Grant No.113F155. E.K. is supported by the TUBITAK 2214-A Scholarship. G.A. acknowledges support by a grant of the Dutch Academy of Sciences (KNAW). We would like to thank M. Gurses, T.C. Sisman, Deniz Olgu Devecioglu, Mehmet Ozkan and Shankhadeep Chakrabortty for useful discussions.


\begin{thebibliography}{0}

\bibitem{BTZ} 
  M.~Banados, C.~Teitelboim and J.~Zanelli,
  ``The Black hole in three-dimensional space-time,''
  Phys.\ Rev.\ Lett.\  {\bf 69}, 1849 (1992).
  

\bibitem{NMG} E.~A.~Bergshoeff, O.~Hohm and P.~K.~Townsend,
  ``Massive Gravity in Three Dimensions,''
  Phys.\ Rev.\ Lett.\  {\bf 102}, 201301 (2009).
  
   \bibitem{KGravity} 
  S.~Deser,
  ``Ghost-free, finite, fourth order D=3 (alas) gravity,''
  Phys.\ Rev.\ Lett.\  {\bf 103}, 101302 (2009).
  
   \bibitem{MMG2} 
  B.~Tekin,
  ``Bulk and boundary unitary gravity in 3D: MMG$_2$,''
  Phys.\ Rev.\ D {\bf 92}, 024008 (2015).

\bibitem{Altas} 
  E.~Altas and B.~Tekin,
  ``Holographically Viable Extensions of Topologically Massive and Minimal Massive Gravity?,''
  arXiv:1512.06651 [hep-th].  
  
   \bibitem{DJT} S.~Deser, R.~Jackiw and S.~Templeton,
  ``Three-Dimensional Massive Gauge Theories,''
  Phys.\ Rev.\ Lett.\  {\bf 48}, 975 (1982).;
  ``Topologically Massive Gauge Theories,''
  Annals Phys.\  {\bf 140}, 372 (1982).
  
  \bibitem{mmg} 
  E.~Bergshoeff, O.~Hohm, W.~Merbis, A.~J.~Routh and P.~K.~Townsend,
  ``Minimal Massive 3D Gravity,''
  Class.\ Quant.\ Grav.\  {\bf 31}, 145008 (2014).
  
  \bibitem{Barnich} 
  G.~Barnich, C.~Troessaert, D.~Tempo and R.~Troncoso,
  ``Asymptotically locally flat spacetimes and dynamical nonspherically-symmetric black holes in three dimensions,''
   Phys.\ Rev.\ D {\bf 93}, 084001 (2016).
 

\bibitem{Tempo} 
  C.~Troessaert, D.~Tempo and R.~Troncoso,
  ``Asymptotically flat black holes and gravitational waves in three-dimensional massive gravity,''
  arXiv:1512.09046 [hep-th].

\bibitem{Fareghbal} 
  R.~Fareghbal and S.~M.~Hosseini,
  ``Holography of 3D Asymptotically Flat Black Holes,''
  Phys.\ Rev.\ D {\bf 91}, no. 8, 084025 (2015).
  
  \bibitem{Oliva} 
  J.~Oliva, D.~Tempo and R.~Troncoso,
  ``Three-dimensional black holes, gravitational solitons, kinks and wormholes for BHT massive gravity,''
  JHEP {\bf 0907}, 011 (2009).
 
 \bibitem{Oliva2} 
  J.~Oliva, D.~Tempo and R.~Troncoso,
  ``Static spherically symmetric solutions for conformal gravity in three dimensions,''
  Int.\ J.\ Mod.\ Phys.\ A {\bf 24}, 1588 (2009).
  
  \bibitem{Blagojevic} 
  M.~Blagojevic and B.~Cvetkovic,
  ``Conformally flat black holes in Poincar\'e gauge theory,''
  arXiv:1510.00069 [gr-qc].
  
\bibitem{Gurses_Gursey} 
  M.~Gurses and G.~Feza,
 ``Lorentz Covariant Treatment of the Kerr-Schild Metric,''
  J.\ Math.\ Phys.\  {\bf 16}, 2385 (1975).

  \bibitem{Gurses1}
M. Gurses, "Perfect fluid sources in 2+1 dimensions"
  Class.\ Quant.\ Grav.\  {\bf 11}, 2585 (1994). 
   
 \bibitem{Gurses2} 
  M.~Gurses,
  ``Killing Vector Fields in Three Dimensions: A Method to Solve Massive Gravity Field Equations,''
  Class.\ Quant.\ Grav.\  {\bf 27}, 205018 (2010)
  [Class.\ Quant.\ Grav.\  {\bf 29}, 059501 (2012)].
  
  \bibitem{gullu} 
  I.~Gullu, T.~C.~Sisman and B.~Tekin,
  ``Born-Infeld extension of new massive gravity,''
  Class.\ Quant.\ Grav.\  {\bf 27}, 162001 (2010).  
  
\bibitem{gullu2} 
  I.~Gullu, T.~C.~Sisman and B.~Tekin,
  ``c-functions in the Born-Infeld extended New Massive Gravity,''
  Phys.\ Rev.\ D {\bf 82}, 024032 (2010).  


\bibitem{Tahsin_tez}
 T. C. Sisman,  Born-Infeld gravity theories in D-dimensions (Doctoral dissertation, PhD thesis, METU 2012).
  


   \bibitem{Nam} 
  S.~Nam, J.~D.~Park and S.~H.~Yi,
  ``AdS Black Hole Solutions in the Extended New Massive Gravity,''
  JHEP {\bf 1007}, 058 (2010).


  \bibitem{Wald}
  J.~Lee and R.~Wald,
  ``Local symmetries and constraints,''
  J.\ Math.\ Phys.\ {\bf 31} 725 (1990).

\bibitem{Wald2}
  R. Wald,
  ``On identically closed forms locally constructed from a field,''
  J.\ Math.\ Phys.\ {\bf 31} 2378 (1990).

\bibitem{Wald3}
  R.~M.~Wald,
  ``Black hole entropy is the Noether charge,''
  Phys.\ Rev.\  D {\bf 48} 3427 (1993).

\bibitem{Jacobson}
  T.~Jacobson, G.~Kang and R.~C.~Myers,
  ``On Black Hole Entropy,''
  Phys.\ Rev.\  D {\bf 49} 6587 (1994).
 
\bibitem{Myers}
  T.~Jacobson, G.~Kang and R.~Myers,
  ``Black hole entropy in higher curvature gravity,''
  [arXiv:gr-qc/9502009].

\bibitem{Saida}
  H.~Saida and J.~Soda,
  ``Statistical entropy of BTZ
  black hole in higher curvature gravity,''
  Phys.\ Lett.\  B {\bf 471}358 (2000) .

  \bibitem{CSentropy} 
  S.~N.~Solodukhin,
  ``Holography with gravitational Chern-Simons,''
  Phys.\ Rev.\ D {\bf 74}, 024015 (2006).
  
 \bibitem{CSentropy2} 
  Y.~Tachikawa,
  ``Black hole entropy in the presence of Chern-Simons terms,''
  Class.\ Quant.\ Grav.\  {\bf 24}, 737 (2007).

\bibitem{Giribet} 
  G.~Giribet, J.~Oliva, D.~Tempo and R.~Troncoso,
  ``Microscopic entropy of the three-dimensional rotating black hole of BHT massive gravity,''
  Phys.\ Rev.\ D {\bf 80}, 124046 (2009).  

  \bibitem{Ahmedovprl} 
  H.~Ahmedov and A.~N.~Aliev,
  ``Exact Solutions in $D=3$ New Massive Gravity,''
  Phys.\ Rev.\ Lett.\  {\bf 106}, 021301 (2011).
  
  \bibitem{Ahmedovplb} 
  H.~Ahmedov and A.~N.~Aliev,
  ``The General Type N Solution of New Massive Gravity,''
  Phys.\ Lett.\ B {\bf 694}, 143 (2011).
  
\bibitem{Gurses:2011fv}
  M.~Gurses, T.~C.~Sisman and B.~Tekin,
  ``Some exact solutions of all $f(R_{\mu\nu})$ theories in three dimensions,''
  Phys.\ Rev.\ D {\bf 86} 024001  (2012).

  \bibitem{Ahmedovprd} 
  H.~Ahmedov and A.~N.~Aliev,
  ``Type D Solutions of $3D$ New Massive Gravity,''
  Phys.\ Rev.\ D {\bf 83}, 084032 (2011).
  

  \end{thebibliography}
\end{document}